\documentclass[amsmath,amssymb,twocolumn]{revtex4}
\usepackage{graphicx}
\usepackage{amscd,amsmath,amsthm,amsfonts,amssymb}

\def\ri{{\rm i}}

\def\re{{\rm e}}

\begin{document}
\title{Pyramid diffraction in parity-time-symmetric optical lattices}
\author{Sean Nixon and Jianke Yang}

\address{Department of Mathematics and Statistics, University of Vermont, Burlington, VT 05401, USA}

\begin{abstract}
Nonlinear dynamics of wave packets in two-dimensional
parity-time-symmetric optical lattices near the phase-transition
point are analytically studied. A novel fourth-order equation is
derived for the envelope of these wave packets. A pyramid
diffraction pattern is demonstrated in both the linear and nonlinear
regimes. Blow-up is also possible in the nonlinear regime for both
focusing and defocusing nonlinearities.
\end{abstract}
\maketitle

Parity-time ($\mathcal{PT}$)-symmetric wave systems have the
unintuitive property that their linear spectrum can be completely
real even though they contain gain and loss \cite{Bender1998}. In
spatial optics, $\mathcal{PT}$-symmetric systems can be realized by
employing symmetric index guiding and an antisymmetric gain/loss
profile \cite{PT_2005,Christodoulides2007,Guo2009,Segev2010}. In
temporal optics, $\mathcal{PT}$-symmetric systems can be obtained as
well \cite{PT_lattice_exp,coupler1,coupler2}. So far, a number of
novel phenomena in optical $\mathcal{PT}$ systems have been
reported, including phase transition, nonreciprocal Bloch
oscillation, unidirectional propagation, distinct pattern of
diffraction, formation of solitons and breathers, wave blowup, and
so on \cite{Guo2009,Segev2010,PT_lattice_exp,coupler1,coupler2,
Musslimani2008,Longhi_2009,Musslimani_diffraction_2010,Christodoulides_uni_2011,Nixon2012,Nixon2012b}.

In this Letter, we analytically study nonlinear dynamics of wave
packets in two-dimensional $\mathcal{PT}$-symmetric optical lattices
near the phase-transition point (where diffraction surfaces of Bloch
bands cross like the intersection of four planes). Near these
intersections we show that the evolution of wave packets is governed
by a novel fourth-order equation. Based on this envelope equation,
we predict a pyramid (i.e., expanding square) diffraction pattern in
both linear and nonlinear regimes. Further, in the nonlinear regime
blow-up can occur for both focusing and defocusing nonlinearities.
These predictions are verified in the full equation as well.

The model for nonlinear propagation of light beams in
$\mathcal{PT}$-symmetric optical lattices is taken as
\begin{equation}
\ri \Psi_z + \nabla^2 \Psi + V(x,y)\Psi + \sigma |\Psi|^2 \Psi = 0,
\label{Eq:NLS}
\end{equation}
where $z$ is the propagation direction, $(x,y)$ is the transverse
plane, $\nabla^2=\partial_{x}^2+\partial_{y}^2$, and $\sigma=\pm 1$
is the sign of nonlinearity. The $\mathcal{PT}$-symmetric potential
$V(x,y)$ is taken as $V(x,y)= \widetilde{V}(x) + \widetilde{V}(y)$,
where
\begin{equation} \label{Eq:Potential1} \widetilde{V}(x) = V_0^2 \left[ \cos(2x)+ \ri W_0
\sin(2x) \right],
\end{equation}
$V_0^2$ is the potential depth and $W_0$ is the relative gain/loss
strength.

We begin by considering the linear diffraction relation of Eq.
\eqref{Eq:NLS} at the phase-transition point $W_0=1$
\cite{Musslimani2008,Nixon2012}. In this case, the linear equation
of \eqref{Eq:NLS} can be solved exactly \cite{Nixon2012}. The
diffraction relation is $\mu = (k_x + 2m_1)^2 + (k_y + 2m_2)^2$,
where $(k_x, k_y)$ are Bloch wavenumbers in the first Brillouin zone
$-1 \le k_x, k_y \le 1$, and $(m_1,~m_2)$ are any pair of
nonnegative integers. The most complex degeneracies occur at points
$k_x = 0, \pm 1$ and $k_y = 0, \pm 1$, where the diffraction surface
intersects itself four-fold as illustrated in Fig. \ref{Fig:
Dispersion}. If a carrier Bloch wave is chosen at one of these
degeneracies, then the envelope of the resulting wave packet
exhibits novel behavior which we elucidate below.

\begin{figure}
\begin{center}
\includegraphics[height = 1.6in]{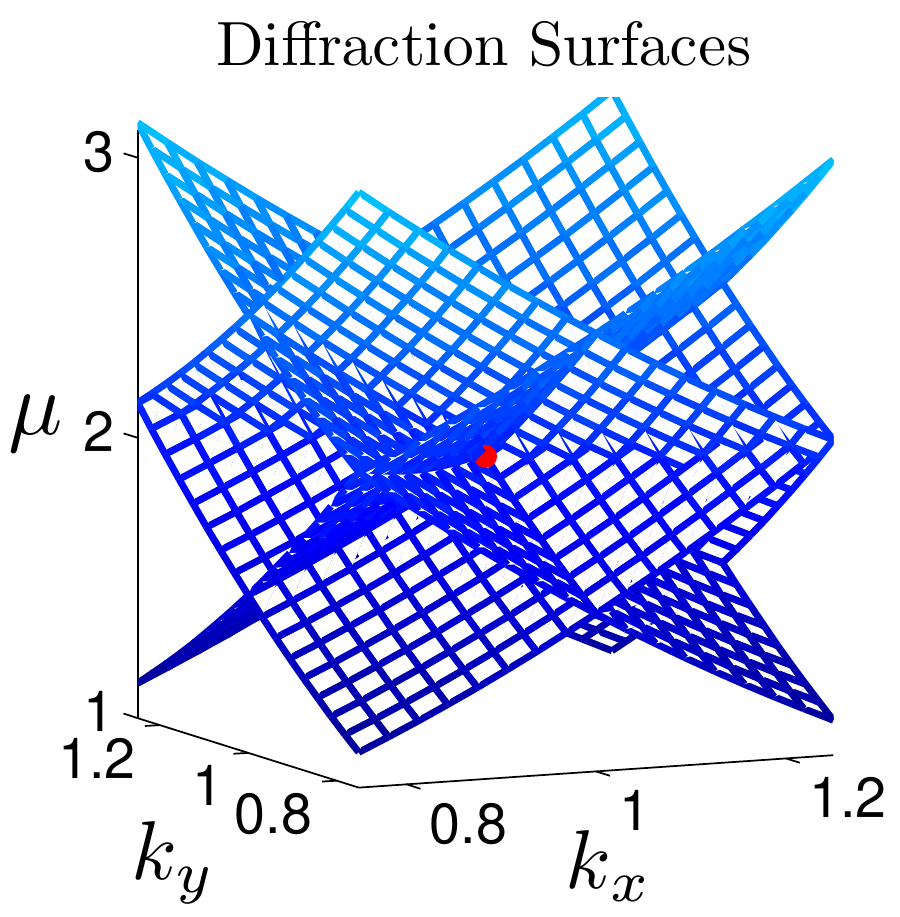} \hspace{0.2cm}
\includegraphics[height = 1.6in]{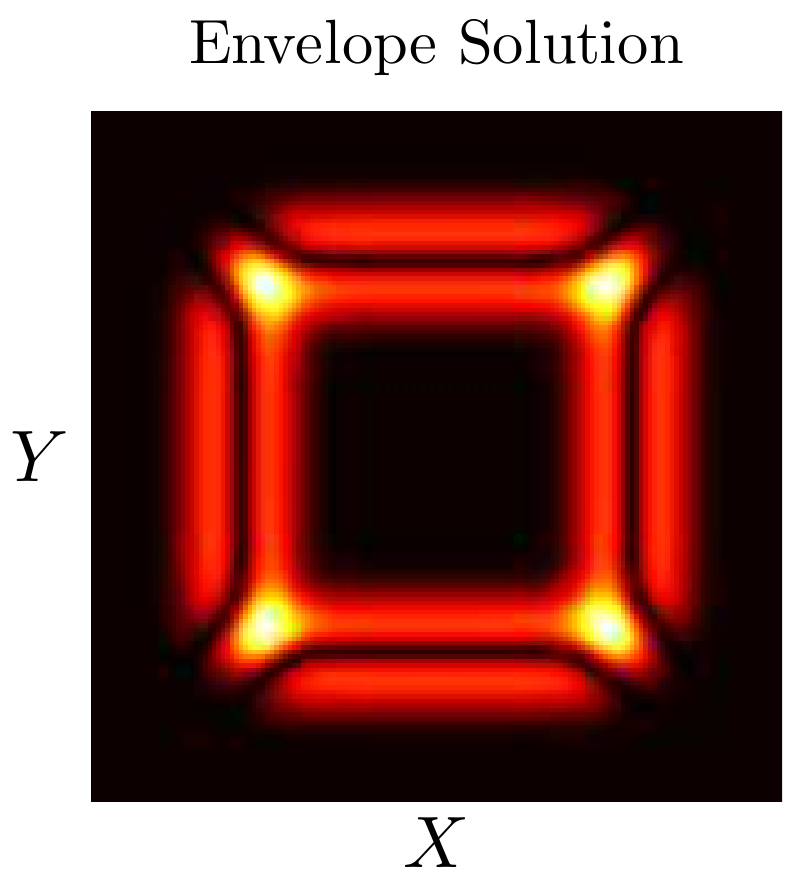}

\vspace{0.3cm}
\includegraphics[height = 1in]{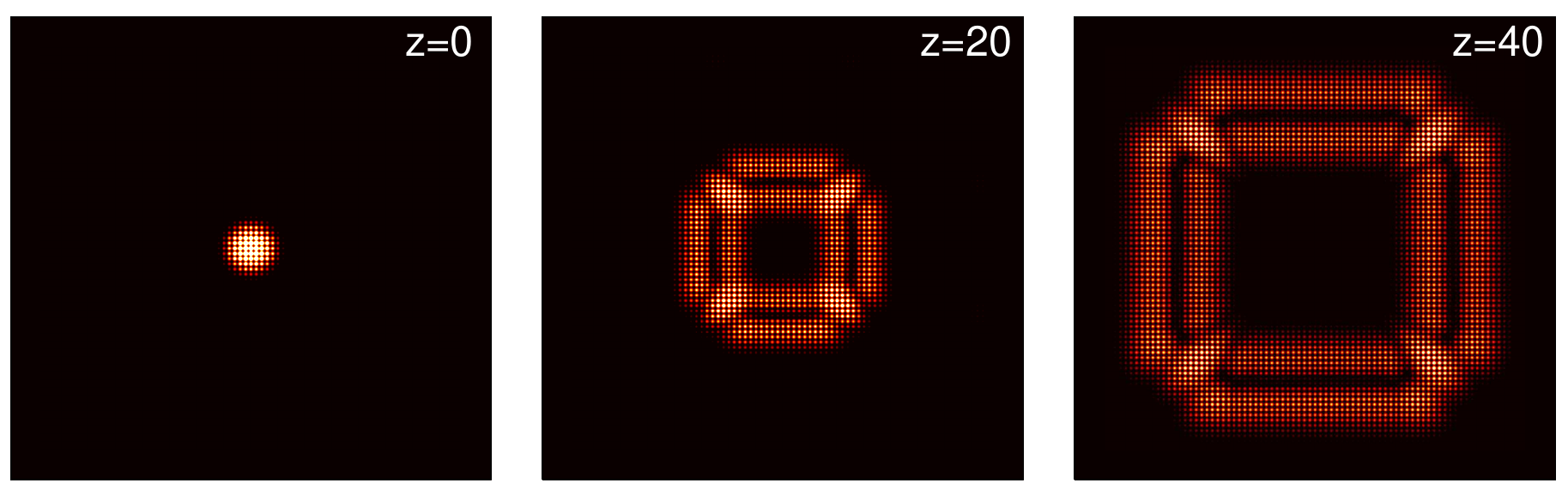}

\caption{(Upper left) Diffraction relation near the intersection
point $(k_x, k_y, \mu) = (1,1,2)$ (marked by a red dot). (Upper
right) Linear-diffraction pattern of an initial Gaussian envelope at
phase transition in the envelope equation (\ref{Eq: EnvelopeA}).
(Lower row) Linear diffraction of an initial Gaussian wave packet at
phase transition in the full equation (\ref{Eq:NLS}). } \label{Fig:
Dispersion}
\end{center}
\end{figure}

To analyze the structure of these degenerices we exploit the fact
that the potential $V(x,y)$ is separable. Posed as an eigenvalue
problem for $\Psi = \phi(x,y) \re^{-\ri \mu z}$ in the linear
equation of \eqref{Eq:NLS}, we get $L \phi  = -\mu \phi$, where $L=
L^{(x)} + L^{(y)}$, $L^{(x)}\equiv \partial_x^2 +
\widetilde{V}_0(x)$, and $\widetilde{V}_0(x)$ is the $\mathcal{PT}$
lattice \eqref{Eq:Potential1} at the phase-transition point $W_0=1$.
At four-fold intersection points, the eigenvalues are $\mu =
n_1^2+n_2^2$, where $(n_1,~n_2)$ are any pair of positive integers.
The operator $L^{(x)}$ ($L^{(y)}$) has eigenvalues $n_1^2$ ($n_2^2$)
with geometric multiplicity 1 and algebraic multiplicity 2
\cite{Nixon2012b}. Let $\phi^{e_1}(x)$ ($\phi^{e_2}(y)$) be the
eigenfunction and $\phi^{g_1}(x)$ ($\phi^{g_2}(y)$) the associated
generalized eigenfunction. Then
\begin{equation}
\phi^{e_1}(x)=\tilde{I}_{n_1}(V_0e^{ix}), \quad
\phi^{e_2}(y)=\tilde{I}_{n_2}(V_0e^{iy}),
\end{equation}
where $\tilde{I}_n(V_0e^{ix})$ is the modified Bessel function
$I_n(V_0e^{ix})$ normalized to have unit peak amplitude, and
$(L^{(x)}+n_1^2)\phi^{g_1}=\phi^{e_1}$, $(L^{(y)}+n_2^2)
\phi^{g_2}=\phi^{e_2}$. Since $L= L^{(x)} + L^{(y)}$, we see that
$L$ has two eigenfunctions
\begin{subequations}\begin{align}
\phi^{01}(x,y) &= \phi^{e_1}(x) \phi^{e_2}(y), \\
\phi^{02}(x,y) &= \phi^{e_1}(x)\phi^{g_2}(y) -
\phi^{g_1}(x)\phi^{e_2}(y).
\end{align}\end{subequations}
In addition, the first eigenfunction $\phi^{01}$ has two generalized
eigenfunctions
\begin{subequations}\begin{align}
\phi^{11}(x,y) &= [\phi^{e_1}(x) \phi^{g_2}(y) + \phi^{g_1}(x) \phi^{e_2}(y)]/2,\\
\phi^{21}(x,y) &= \phi^{g_1}(x) \phi^{g_2}(y),
\end{align}\end{subequations}
where $(L+\mu) \phi^{11} = \phi^{01}$, and $(L+\mu) \phi^{21} =
\phi^{11}$.

We now study the nonlinear dynamics of wave packets near these
intersections. For simplicity, we will conduct the analysis at the
lowest intersection point, $\mu =2$. But similar results can be
obtained for any intersection point.

\begin{figure}
\begin{center}
\includegraphics[height = 1in]{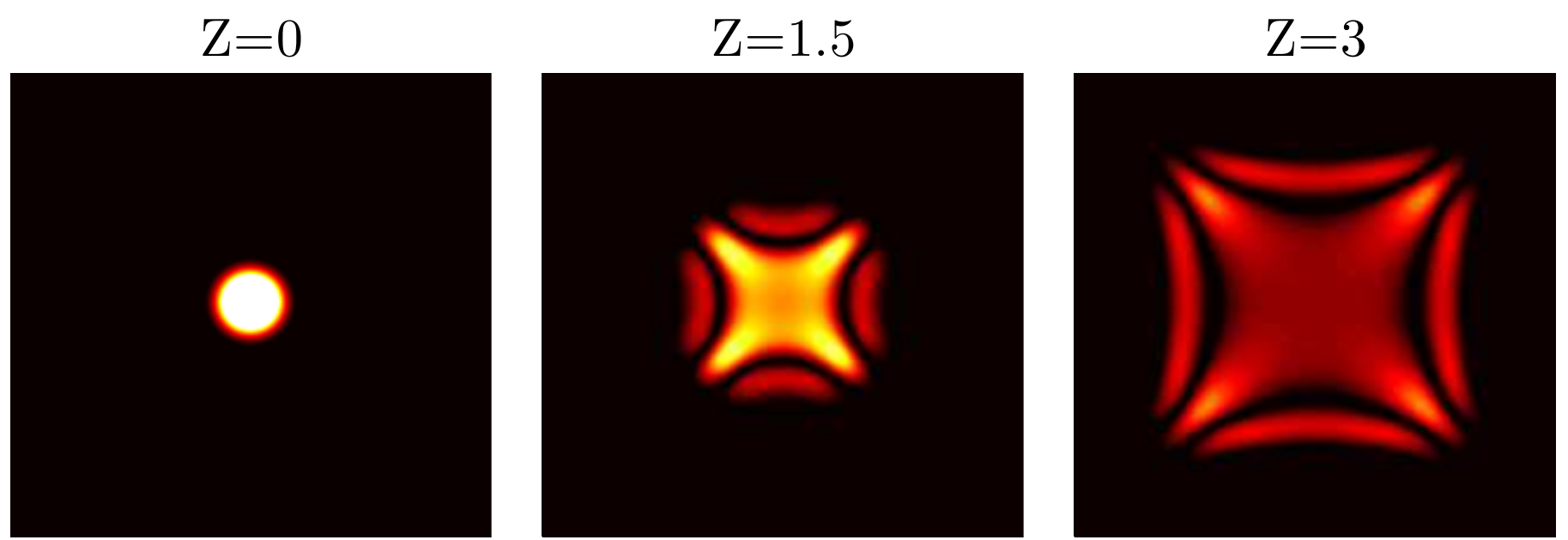}

\includegraphics[height = 1in]{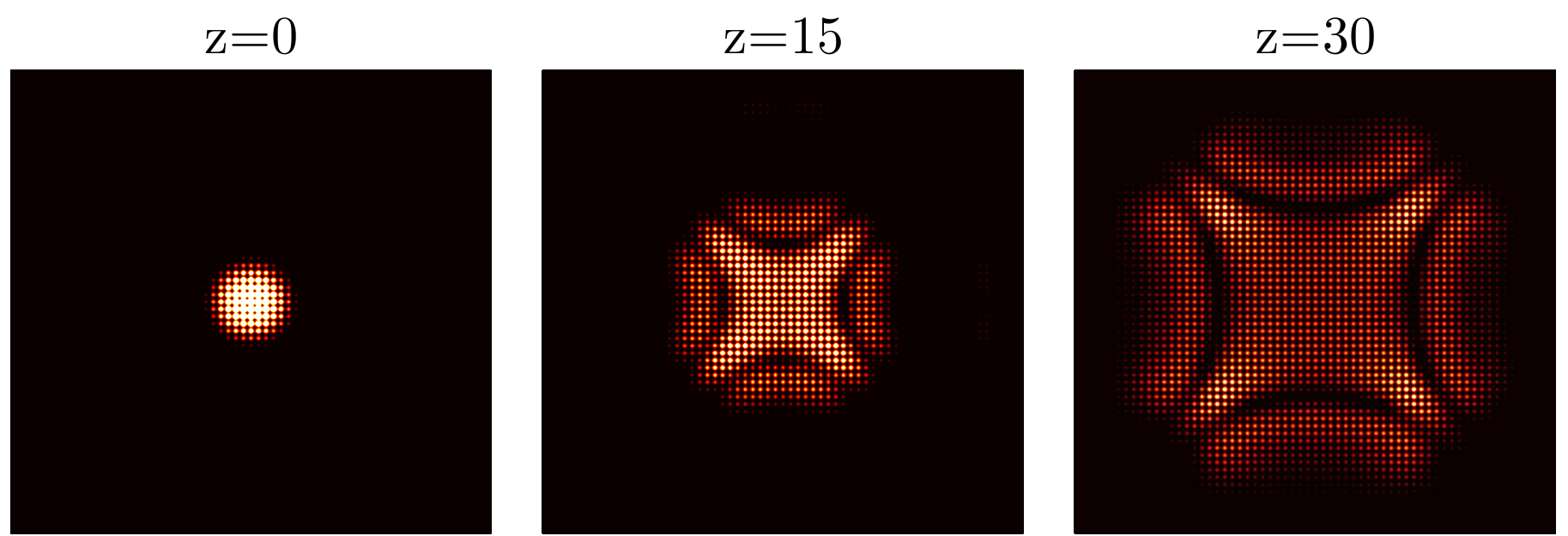}
\caption{Pyramid diffraction of a Gaussian wavepacket in the linear
equation below phase transition. Upper row: diffraction in the
envelope equation (\ref{Eq: EnvelopeA}); lower row: diffraction in
the full equation (\ref{Eq:NLS}). } \label{Fig: LinearEvolve}
\end{center}
\end{figure}

The perturbation expansion for the wave packet near the intersection
$\mu=2$ is
\begin{equation}
\Psi = \epsilon^{\frac{3}{2}}\psi \re^{-\ri \mu t}, \quad  \psi=
\psi_0  + \epsilon \psi_1  + \ldots, \label{Eq: Expansion}
\end{equation}
where $\psi_0 = A(X,Y,Z) \phi^{01}(x,y)$ is the leading-order wave
packet for the Bloch mode $\phi^{01}$ at the intersection point
$(k_x, k_y, \mu)=(1,1, 2)$, $(X,Y,Z) =(\epsilon x, \epsilon y,
\epsilon z)$ are slow spatial variables, and $0< \epsilon \ll 1$.
Near the phase-transition point, $W_0$ can be expressed as $W_0 = 1
- \eta \epsilon^2/V_0^2$, where $\eta$ measures the deviation from
phase-transition. After introducing the slow variables into equation
\eqref{Eq:NLS}, the new equation for $\psi$ is
\begin{align}
\label{Eq: SlowVariables} (L+ \mu) & \psi  =  - \ri \epsilon \psi_Z
- 2\epsilon (\psi_{xX}+\psi_{yY})
 - \epsilon^2(\partial_X^2+\partial_Y^2)\psi  \hspace{0.5cm} \notag
 \\ & +\ri \eta \epsilon^2  [\sin(2x)+ \sin(2y)] \psi - \epsilon^3 \sigma |\psi|^2
 \psi.
\end{align}

We proceed by inserting expansion \eqref{Eq: Expansion} into
equation \eqref{Eq: SlowVariables} and solving for $\psi_n$ at each
order. Each $\psi_n$ satisfies a linear inhomogeneous equation, with
the homogeneous operator being $L+\mu$. In order for it to be
solvable, the Fredholm conditions need to be satisfied, i.e., the
inhomogeneous term must be orthogonal to the kernels $\phi^{01*}$
and $\phi^{02*}$ of the adjoint operator $L^*+\mu$. Here $*$ stands
for complex conjugation.

At O$(\epsilon)$ the solvability conditions for $\psi_1$ are
automatically satisfied, and thus we can solve $\psi_1$ as
\begin{equation}
\psi_1 = -\ri A_T \phi^{11} - 2A_X \phi^{a} - 2 A_Y \phi^{b} + B
\phi^{02},
\end{equation}
where $B(X,Y,Z)$ is the envelope of the second eigenfunction
$\phi^{02}$, $(L+\mu) \phi^{a} = \phi^{01}_x$, $(L+\mu) \phi^{b} =
\phi^{01}_y$, and $\phi^{a}$, $\phi^{b}$ are assumed to be
orthogonal to $\phi^{01}$ and $\phi^{02}$.

Now we proceed to the $\psi_2$ equation at O$(\epsilon^2)$. The
orthogonality condition with $\phi^{01*}$ is automatically
satisfied, and the orthogonality condition with $\phi^{02*}$ gives
\begin{equation}
\ri B_Z = A_{ZX} - A_{ZY} + 2A_{XX} -2A_{YY},  \label{Eq: BZ}
\end{equation}
which defines the connection between envelopes $A$ and $B$ of the
two eigenmodes at the Bloch-surface intersection. Under this
relation, the $\psi_2$ equation can be solved.

Finally we proceed to the $\psi_3$ equation at O$(\epsilon^3)$. The
orthogonality condition with $\phi^{01*}$ gives
\begin{eqnarray} \label{Eq: AZZZ}
& \partial_Z^3 A - 8(\partial_X^2 +\partial_Y^2) \partial_Z A -
8(\partial_X^2 - \partial_Y^2)(\partial_X - \partial_Y)A
\hspace{0.3cm} \nonumber
\\
& \hspace{-0.8cm} + 8(\partial_X^2 - \partial_Y^2)( \ri  B) + \alpha
\partial_Z A+ \ri \tilde{\sigma} |A|^2A =0,
\end{eqnarray}
where
\[
\alpha=2V_0^2\eta, \qquad \tilde{\sigma}=-\ri \sigma
\frac{\int_0^{2\pi} \int_0^{2\pi}  |\phi^{01}|^2\phi^{01}\phi^{21}
dxdy}{\int_0^{2\pi} \int_0^{2\pi}  \phi^{01}\phi^{21} dxdy}.
\]
This equation, combined with equation \eqref{Eq: BZ}, yields a
single fourth-order envelope equation for $A$ as
\begin{align} \label{Eq: EnvelopeA}
\partial_Z^4 A &- 8 (\partial_X^2 + \partial_Y^2) \partial_Z^2  A + 16 (\partial_X^2-\partial_Y^2)^2A  \notag \\
& + \alpha \partial_Z^2A+\ri\tilde{\sigma}  \partial_Z \left(|A|^2A
\right) =0.
\end{align}
This novel envelope equation is one of the main results in this
Letter.

It remains to relate the initial conditions for $\Psi$ with those
for the envelope equation (\ref{Eq: EnvelopeA}). By collecting the
$\psi_0, \psi_1$ and $\psi_2$ solutions from the above analysis and
projecting the resulting perturbation-series solution (\ref{Eq:
Expansion}) onto the eigenfunctions and generalized eigenfunctions,
we find the dominant terms of $\Psi$ are given by
\begin{equation}\label{Eq: psiABCD}
\Psi \approx \epsilon^{\frac{3}{2}} (A \phi^{01} + \epsilon B
\phi^{02}+ \epsilon C \phi^{11}+\epsilon^2 D \phi^{21}),
\end{equation}
where
\begin{equation}\label{Eq: C}
C = -\ri \left( A_Z + 2A_X + 2A_Y \right),
\end{equation}
\begin{equation}\label{Eq: D}
D =- A_{ZZ} - 2A_{ZX} - 2A_{ZY} - \frac{\alpha}{2} A
 + 4\ri B_{X} - 4\ri B_{Y}.
\end{equation}
Thus, from initial envelope functions $A,B,C,D$ of the
eigenfunctions and generalized eigenfunctions in $\Psi$, we can
obtain initial conditions for $A, A_Z, A_{ZZ}$ and $A_{ZZZ}$ from
(\ref{Eq: AZZZ}), (\ref{Eq: C}) and (\ref{Eq: D}).


Direct simulations show strong agreement between the envelope
dynamics in \eqref{Eq: EnvelopeA} and those corresponding wave
packets in the full equation \eqref{Eq:NLS}. For this letter we take
the initial conditions
\begin{equation} \label{Eq:ic}
A= A_0\re^{-(X^2+Y^2)}, \, A_Z = A_{ZZ} = 0, \, \partial_Z^3 A = -
\ri \tilde{\sigma} |A|^2A
\end{equation}
in the envelope equation, or the equivalent initial conditions based
on \eqref{Eq: psiABCD} for simulations of the full equation
\eqref{Eq:NLS}. For the constants we take $V_0^2 = 6$, $\epsilon =
0.1$, and $\eta=0$ or $1$ (at or below phase transition
respectively). Then we find $\alpha = 12\eta$, and
$\tilde{\sigma}\approx 7.3\sigma$.

\begin{figure}
\begin{center}
\includegraphics[height = 1in]{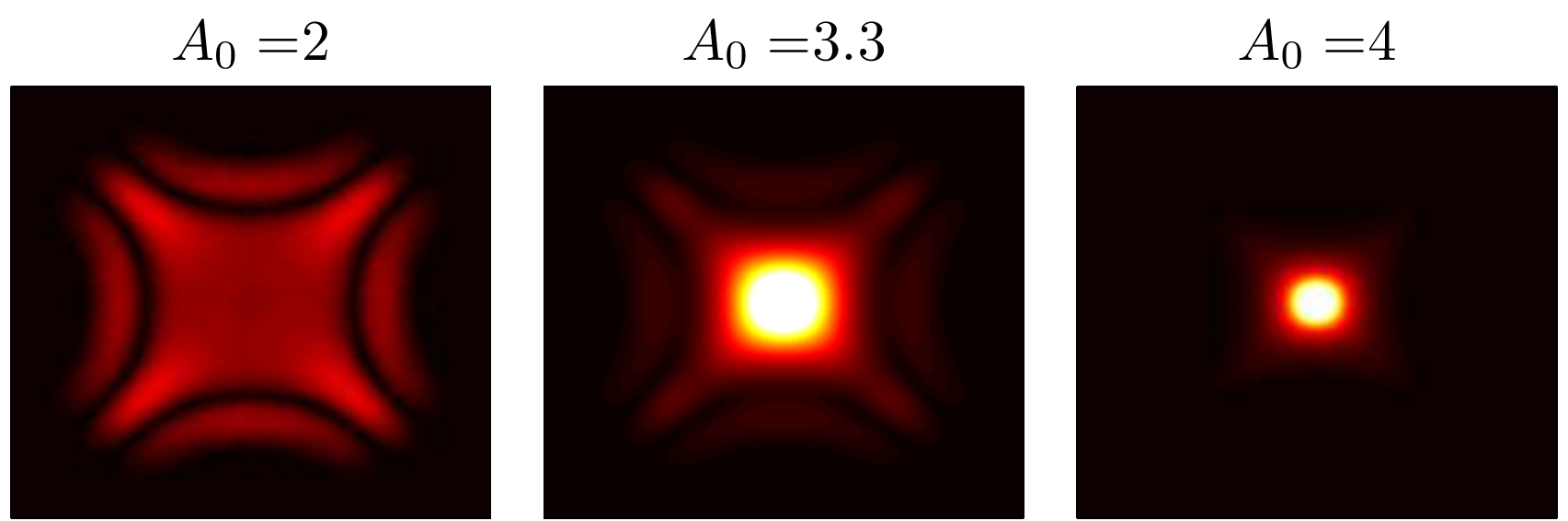}

\vspace{0.15cm}
\includegraphics[height = 1in]{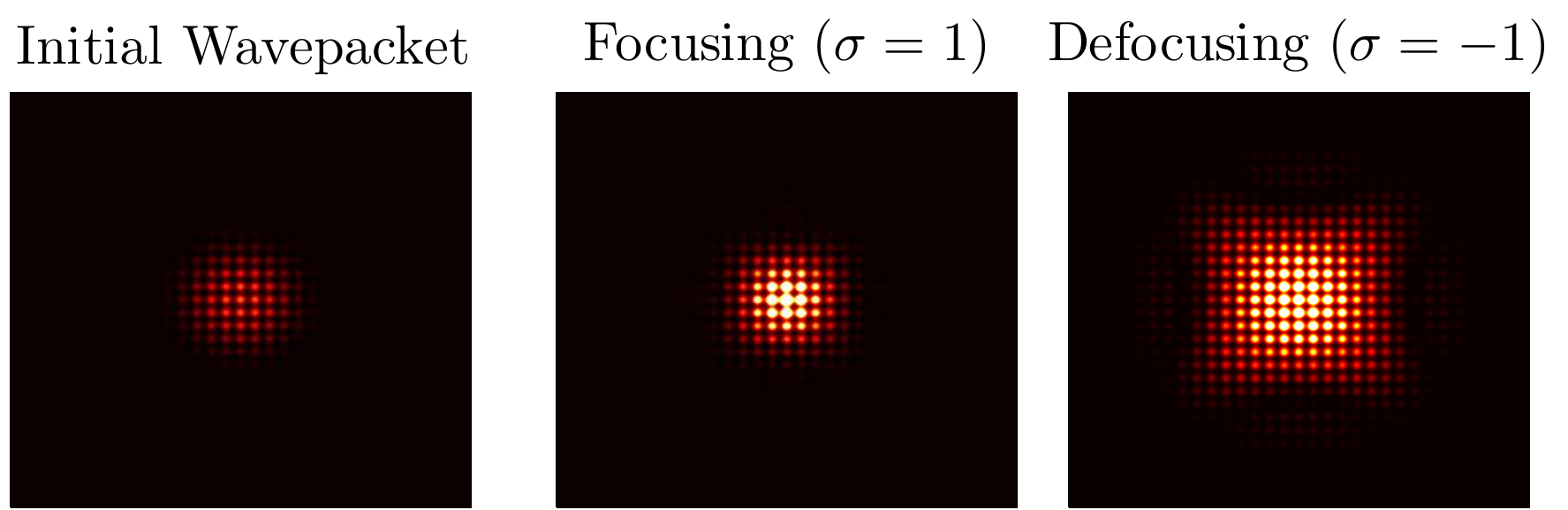}
\caption{Nonlinear dynamics of wave packets below phase transition.
Upper row: envelope solutions in (\ref{Eq: EnvelopeA}) at $Z\approx
2$ for three values of $A_0$ in (\ref{Eq:ic}). Lower row: solutions
of the full equation (\ref{Eq:NLS}) for the initial wavepacket with
$A_0=6$ (left) at later distances under focusing (middle) and
defocusing (right) nonlinearities. } \label{Fig: NonlinearEvolve}
\end{center}
\end{figure}

In the linear equation at the phase transition point, i.e.,
$\alpha=\tilde{\sigma} = 0$, equation \eqref{Eq: EnvelopeA} has the
general solution
\begin{align}\label{Eq: ASolution}
A&= A_1(X-2Z, Y-2Z) + A_2(X-2Z, Y+2Z)  \notag \\
 &+ A_3(X+2Z, Y-2Z) +A_4(X+2Z, Y+2Z)
\end{align}
for arbitrary $A_n$ functions. In general this corresponds to an
expanding square wave front propagating with speeds $\pm 2$ in both
$X$ and $Y$ directions, which we term pyramid diffraction. This
pattern is illustrated in Fig. \ref{Fig: Dispersion} for both the
envelope and full equations. Notice that the wave fronts are flat on
all four sides.

In the linear equation but below the phase transition point
($\alpha=12$, $\tilde{\sigma}=0$), the pyramid diffraction is
qualitatively similar to that in Fig. 1, with wave fronts expanding
roughly like a square. But the wave fronts are no longer flat. In
addition, the core develops an `x' shape. An example is shown in
Fig. \ref{Fig: LinearEvolve}, where diffractions in both the
envelope and full equations are displayed.

In the presence of nonlinearity ($\tilde{\sigma}\approx 7.3\sigma$)
and below phase transition, the wave packet diffracts away if its
initial amplitude is below a certain threshold value. This nonlinear
diffraction is also pyramid-like, closely resembling the linear
pyramid diffraction in Fig. 2. An example is displayed in Fig. 3
(upper left panel). However, if the initial amplitude is above this
threshold, the envelope solution blows up to infinity in finite
distance. For example, with the initial condition (\ref{Eq:ic}), the
envelope solution in (\ref{Eq: EnvelopeA}) blows up when $A_0>3.2$.
These blowup solutions are displayed in Fig. 3 (upper middle and
right panels). Remarkably, this blowup is independent of the sign of
the nonlinearity, a fact which is clear from the envelope equation
(\ref{Eq: EnvelopeA}), since a sign change in $\tilde{\sigma}$ can
be accounted for by taking the complex conjugate of this equation.
In the full equation (\ref{Eq:NLS}), we have confirmed that similar
growth occurs for both signs of the nonlinearity as well. For
instance, evolution of a wavepacket (corresponding to $A_0=6$) under
focusing and defocusing nonlinearities are displayed in Fig. 3
(lower row). In both cases solutions rise to very high amplitudes as
the envelope equation predicts.

In summary, we have analyzed nonlinear dynamics of wave packets in
two-dimensional $\mathcal{PT}$-symmetric lattices near the
phase-transition point. In the linear regime, pyramid diffraction is
demonstrated. In the nonlinear regime, wave blowup is obtained for
both focusing and defocusing nonlinearities.

This work is supported in part by AFOSR.

\end{document}